\documentclass[english,aps,prd,
groupedaddress,amssymb,
showpacs,nofootinbib]{revtex4}

\usepackage{times}
\usepackage{amsmath}

\newcommand{\beq}{\begin{eqnarray}}
\newcommand{\eeq}{\end{eqnarray}}

\makeatletter
\numberwithin{equation}{section}

\makeatother

\usepackage{babel}

\begin{document}
\setlength{\unitlength}{1mm}

\title{Correlation Functions of the SU($\infty$) Principal Chiral Model}

\author{Axel \surname{Cort\'es Cubero}$^{ab}$}

\email{acortes_cubero@gc.cuny.edu}

\author{Peter \surname{Orland}$^{abc}$}

\email{orland@nbi.dk}

\affiliation{ a. Baruch College, The 
City University of New York, 17 Lexington Avenue, 
New 
York, NY 10010, U.S.A. \\
b. The Graduate School and University Center, The City University of New York, 365 Fifth Avenue,
New York, NY 10016, U.S.A.\\
c. The Niels Bohr Institute, Blegdamsvej 17, DK-2100, Copenhagen \O, Denmark}

\begin{abstract}
We obtain exact matrix elements of physical operators of the ($1+1$)-dimensional nonlinear sigma model of an  SU($N$)-valued bare field, in the 't$\;$Hooft limit $N\rightarrow\infty$. Specifically, all the form factors of the Noether current and the 
stress-energy-momentum tensor are found with an integrable bootstrap method. These form factors are used to find vacuum expectation values of products of these operators.
\end{abstract}

\maketitle

\section{Introduction}

In this paper we find all the form factors of the Noether current and stress-energy-momentum tensor
operators of the principal chiral sigma model in $(1+1)$ dimensions, in the limit of large $N$. This field theory has the action 
\beq
S=\frac{N}{2g_{0}^{2}}\int d^{2}x\;\eta^{\mu\nu}{\rm Tr}\,\partial_\mu U(x)^\dagger\partial_\nu U(x),
\label{action}
\eeq
where the field $U(x)$ is in the fundamental representation of SU($N$), $\mu,\nu=0,1,$ and 
$\eta^{00}=1, \eta^{11}=-1, \eta^{10}=\eta^{01}=0$. The action is unchanged by a 
global ${\rm SU}(N)\!\times \!{\rm SU}(N)$ transformation, $U(x)\to V_L U(x) V_R$, where 
$V_L, V_R\in {\rm SU}(N)$. This model is asymptotically free and has a 
mass gap, which we denote by $m$. It is also completely integrable
and its S matrix is known \cite{Wiegmann}. The large-$N$ limit we consider is the 't$\;$Hooft limit of $N\rightarrow \infty$, with $g_{0}^{2}$ fixed. We assume that the mass gap is fixed in this limit. We do not consider alternative large-$N$ limits, such as that of Fateev {\em et. al.} \cite{FKW}, in which the mass gap vanishes. For reviews of the large-$N$ limit
of unitary-matrix models, see References \cite{RCV}.

Recently, the integrable bootstrap method was used to calculate all the form factors of the renormalized-field  operator $\Phi(x)$ at large $N$ \cite{Orland1}, \cite{Orland2}. 
This was used to find an exact expression for the Wightman function, {\em i.e.}, the non-time-ordered vacuum expectation value of two renormalized field operators
\beq
W^{\Phi}(x)=\frac{1}{N}\langle0\vert{\rm Tr}\,\Phi(x)\Phi(0)^\dagger\vert 0\rangle=\lim_{\Lambda \rightarrow \infty}\frac{1}{N}Z[g_{0}(\Lambda),\Lambda]\;\langle0\vert{\rm Tr}\,U(x)U(0)^\dagger\vert 0\rangle,\nonumber
\eeq
where $Z[g_{0}(\Lambda),\Lambda]$ is the field renormalization factor, and $\Lambda$ is an ultraviolet momentum cutoff. The two- and four-excitation form factors of the Noether current operators: 
\beq
j_\mu^L(x)^c_a=-\frac{{\rm i}N}{2g_{0}^{2}}\partial_\mu U_{ab}(x) U^{*cb}(x),\,\,\,
j_\mu^R(x)^d_b=-\frac{{\rm i}N}{2g_{0}^{2}}U^{*ad}(x)\partial_\mu U_{ab}(x),\nonumber
\eeq
where $a,b=1,...,N$ (associated with the symmetries $U\to V_LU$ and $U\to UV_R$, respectively), were also found \cite{Axel}. 

Another operator of interest is the stress-energy-momentum tensor:
\beq
T_{\mu\nu}(x)=\frac{1}{2\pi}\left(\delta^\alpha_\mu\delta_\nu^\beta+\delta_\mu^\beta\delta_\nu^\alpha-\eta_{\mu \nu}\eta^{\alpha \beta}\right)
{\rm Tr}\,\partial_{\alpha}U(x)^{\dagger} \partial_{\beta} U(x)+\lambda \eta_{\mu \nu}, \label{defset}
\eeq
where $\lambda$ is chosen to normal order this operator, so that the vacuum energy is implicitly zero.
There is less freedom 
to define a stress-energy-momentum tensor quadratic in derivatives than for ordinary scalar field theories 
\cite{Coleman}, because the bare field is a unitary matrix. The form of the three terms in parentheses in
(\ref{defset}) are fixed by energy conservation. There is no color-singlet total divergence of dimension
two we can add to the right-hand side. Therefore, it seems that (\ref{defset}) is the most general renormalizable
operator we can define.

The non-time-ordered correlation function of two operators $\mathcal O_{1}(x)$ and $\mathcal O_{2}(0)$ is 
\beq
W^{1,2}_{\mu\nu}(x)=\frac{1}{N}\langle0 \vert
\mathcal O_{1}(x)\mathcal O_{2}(0)
\vert 0\rangle
=\frac{1}{N}\sum_{\Psi}\langle0\vert \mathcal O_{1}(x) \vert \Psi\rangle_{\rm in}\;_{\rm in}\langle\Psi \vert  \mathcal O_{2}(0)\vert 0\rangle, \nonumber
\eeq
where $\langle0\vert\mathcal O_{j}(x)\vert\Psi\rangle_{\rm in}$ is a form factor. Smirnov's axioms
\cite{Smirnov} (see also References \cite{Babujian}) are an effective framework for determining form factors in integrable quantum field theories.

Finding the form factors of  
non-Abelian models with bound states is technically quite delicate. An exception is the matrix element of the current between the vacuum and a particle-antiparticle state, first done for $N=2$ (this is the O($4$) nonlinear sigma model) \cite{KarowskiWeisz}. No bound state can form in this channel, making it possible to evaluate this matrix element for any $N$ \cite{Axel}. The large-$N$ result is confirmed by taking the 
limit. Some progress has been made for the SU($N$) chiral Gross-Neveu model \cite{BabujianKarowEtc.}. It has recently been pointed out \cite{BrittonFrolov}
that the latter model has a free-field representation for the Zamolodchikov-Faddeev algebra \cite{Lukyanov}. The $N=2$ case of the principal chiral model is also known to have 
a free-field representation \cite{FatLashk}. 

Regardless of whether the finite $N$ problem can be generally solved, there are compelling reasons for studying the 
$N\rightarrow\infty$ limit. These include: \begin{enumerate}
\item All the form factors can be found, which makes exact expressions for correlation functions possible.
\item Some of the conventional wisdom concerning the 't$\;$Hooft limit can be tested. For example, the operators defining the Zamolodchikov-Faddeev algebra have been identified with a free Gaussian
master field \cite{Orland1}, from which the scaling field and other operators can be constructed.
\item If form factors are eventually found for any finite $N$, they can be compared to our results by taking
$N\rightarrow \infty$. 
\item There is a striking simplification of the commutation relations of operators, when applying Watson's theorem in the planar limit. This suggests an 
extension to non-integrable field theories in the 't$\;$Hooft limit (we say more about this in Section VII).
\end{enumerate}

This paper is not an introduction to the form-factor program at large $N$, but should be accessible to readers who understand the main points in References 
\cite{Orland1}, \cite{Orland2} and \cite{Axel}.

In the next section, we review a few general facts concerning the principal chiral model. In Section III, we build on earlier results \cite{Orland1}, \cite{Orland2}, \cite{Axel} to find all the form factors of the current operator. We use these
to write down an expression for the vacuum expectation value of two currents in Section IV. We find the form factors of the stress-energy-momentum tensor in Section V; we present the vacuum expectation value of the product
of two such tensors in Section VI. We discuss the effective Abelian operator algebra in Section VII. In the last section, we present some conclusions and directions for future investigation.

\section{The Principal Chiral model}

The principal chiral sigma model has elementary particles and antiparticles with mass $m$. These 
form $r$-particle bound states of mass \cite{Wiegmann}.
\beq
m_r=m\frac{\sin\left(\frac{\pi r}{N}\right)}{\sin\left(\frac{\pi}{N}\right)},\,r=1,\dots,N-1.\nonumber
\eeq
A sine-law bound-state mass spectrum is a general feature of any integrable theory with an attractive interaction and one fundamental 
particle \cite{Schroer}. In the planar limit, $N\rightarrow \infty$, with $m>0$ fixed and $m_{r}$ finite, the binding energy vanishes, except for 
$r=N-1$. Therefore the only remaining bound state is the antiparticle. We describe asymptotic states using the term ``excitation" instead of ``particle", because we wish to consistently distinguish particles from antiparticles.

We create particles and antiparticles by acting with creation operators $\mathfrak{A}^\dagger_P(\theta)_{ab}$ and $\mathfrak{A}^\dagger_A(\theta)_{ba}$, respectively, on the vacuum, where $\theta$ is the rapidity, defined in terms of the momentum vector by $p_{0}=m\cosh\theta,\,p_{1}=m\sinh\theta$, and $a,b=1,2,\dots,N$ are left and right color indices, respectively.

A multiexcitation in-state is made by acting on the vacuum state with a product of creation operators in the order of increasing rapidity, from left to right,
\beq
\vert P,\theta_{1},a_{1},b_{1};A,\theta_{2},b_{2},a_{2};\dots\rangle_{\rm in}=\mathfrak{A}^\dagger_P(\theta_{1})_{a_{1}b_{1}}\mathfrak{A}^\dagger_A(\theta_{2})_{b_{2}a_{2}}\dots\vert 0\rangle,\,{\rm where}\,\theta_{1}>\theta_{2}>\dots\,.\nonumber
\eeq
The two-particle S matrix, $S_{PP}$, defined by
\beq
\,_{\rm out}\langle P,\theta^{\prime}_{1},c_{1},d_{1};P,\theta^{\prime}_{2},c_{2},d_{2}\vert P,\theta_{1},a_{1},b_{1};P,\theta_{2},a_{2},b_{2}\rangle_{\rm in}=S_{PP}(\theta)^{c_{2}d_{2};c_{1}d_{1}}_{a_{1}b_{1};a_{2}b_{2}}4\pi\delta(\theta_{1}-\theta^{\prime}_{1})4\pi\delta(\theta_{2}-\theta^{\prime}_{2}),\nonumber
\eeq
is \cite{Wiegmann}, 
\beq
S_{PP}(\theta)^{c_{2}d_{2};c_{1}d_{1}}_{a_{1}b_{1};a_{2}b_{2}}=S(\theta,N)\left(\delta^{c_{1}}_{a_{1}}\delta^{c_{2}}_{a_{2}}-\frac{2\pi {\rm i}}{N\theta}\delta^{c_{1}}_{a_{2}}\delta^{c_{2}}_{a_{1}}\right)\left(\delta^{d_{1}}_{b_{1}}\delta^{d_{2}}_{b_{2}}-\frac{2\pi {\rm i}}{N\theta}\delta^{d_{1}}_{b_{2}}\delta^{d_{2}}_{b_{1}}\right),\nonumber
\eeq
where
\beq
S(\theta,N)=\frac{\sinh\left(\frac{\theta}{2}-\frac{\pi {\rm i}}{N}\right)}{\sinh\left(\frac{\theta}{2}+\frac{\pi {\rm i}}{N}\right)}\left[\frac{\Gamma(i\theta/2\pi +1)\Gamma(-i\theta/2\pi-1/N)}{\Gamma(i\theta/2\pi+1-1/N)\Gamma(-i\theta/2\pi)}\right]^{2}=1+\mathcal{O}\left(\frac{1}{N^{2}}\right),\nonumber
\eeq
and $\theta=\theta_{1}-\theta_{2}$ is the rapidity difference.

The antiparticle-particle S matrix, $S_{AP}$, is related to the particle-particle S matrix by crossing, $\theta\to\hat{\theta}=\pi {\rm i}-\theta$:
\beq
S_{AP}(\theta)^{d_{2} c_{2}; c_{1} d_{1}}_{a_{1}b_{1};b_{2} a_{2}}=S(\hat{\theta},N)\left(\delta^{c_{1}}_{a_{1}}\delta^{c_{2}}_{a_{2}}-\frac{2\pi {\rm i}}{N\hat{\theta}}\delta_{a_{1}a_{2}}\delta^{c_{1} c_{2}}\right)\left(\delta^{d_{1}}_{b_{1}}\delta^{d_{2}}_{b_{2}}-\frac{2\pi {\rm i}}{N\hat{\theta}}\delta_{b_{1}b_{2}}\delta^{d_{1}d_{2}}\right).\nonumber
\eeq

The creation operators of excitations satisfy the Zamolodchikov algebra:
\beq
\mathfrak{A}^\dagger_P(\theta_{1})_{a_{1}b_{1}}\mathfrak{A}^\dagger_P(\theta_{2})_{a_{2} b_{2}}&=&S_{PP}(\theta)^{c_{2} d_{2} ; c_{1} d_{1}}_{a_{1} b_{1};a_{2} b_{2}}\mathfrak{A}^\dagger_P(\theta_{2})_{c_{2}d_{2}}\mathfrak{A}^\dagger_P(\theta_{1})_{c_{1}d_{1}},\nonumber\\
\mathfrak{A}^\dagger_A(\theta_{1})_{b_{1}a_{1}}\mathfrak{A}^\dagger_A(\theta_{2})_{b_{2} a_{2}}&=&S_{AA}(\theta)^{d_{2} c_{2} ; d_{1} c_{1}}_{b_{1}a_{1};b_{2} a_{2}}\mathfrak{A}^\dagger_A(\theta_{2})_{d_{2}c_{2}}\mathfrak{A}^\dagger_A(\theta_{1})_{d_{1}c_{1}},\nonumber\\
\mathfrak{A}^\dagger_P(\theta_{1})_{a_{1}b_{1}}\mathfrak{A}^\dagger_A(\theta_{2})_{b_{2} a_{2}}&=&S_{AP}(\theta)^{d_{2} c_{2}; c_{1} d_{1}}_{a_{1} b_{1};b_{2} a_{2}}\mathfrak{A}^\dagger_A(\theta_{2})_{d_{2} c_{2}}\mathfrak{A}^\dagger_P(\theta_{1})_{c_{1}d_{1}}.\label{zamoalgebra}
\eeq

\section{General form factors of the current operator}

Under a global ${\rm SU}(N)\! \times \! {\rm SU}(N)$ transformation, the current and particle-creation operators transform as
\beq
j_\mu^L(x)\to V_Lj_\mu^L(x)V^\dagger_L,\,\mathfrak{A}^\dagger_P(\theta)\to V^\dagger_R\mathfrak{A}^\dagger_P(\theta)V^\dagger_L,\,\mathfrak{A}^\dagger_A(\theta)\to V_L\mathfrak{A}^\dagger_A(\theta)V_R.\nonumber
\eeq
Consequently, only form factors with the same number of particles and antiparticles do not vanish. We will call this number $M$, so that the total number of excitations is $2M$.

The $M=1$ and $M=2$ form factors
are \cite{Axel} 
\beq
\langle 0\vert j_\mu^L(x)_{a_{0}c_{0}}\!\!\!&\vert&\!\!\!A,\theta_{1},b_{1},a_{1};P,\theta_{2},a_{2},b_{2}\rangle_{\rm in}\nonumber\\
&=&2\pi {\rm i}\;(p_{1}-p_{2})_\mu\;\frac{\delta_{b_{1}b_{2}}}{\theta_{12}+\pi {\rm i}}\left(\delta_{a_{0}a_{2}}\delta_{c_{0}a_{1}}-\frac{1}{N}\delta_{a_{0}c_{0}}\delta_{a_{1}a_{2}}\right)
 e^{-ix\cdot(p_{1}+p_{2})}+O\left(\frac{1}{N^{2}}\right),
\label{twoparticle}
\eeq
and
\beq
\langle 0\vert j_\mu^L(x)_{a_{0}c_{0}}\!\!\!&\vert&\!\!\!A,\theta_{1}, b_{1}, a_{1};A,\theta_{2},b_{2}a_{2};P,\theta_{3},a_{2},b_{3};P,\theta_{4},a_{4},b_{4}
\rangle_{\rm in}\nonumber\\
&=&\frac{8\pi^{2} i}{N}\left(p_{1}+p_{2}-p_{3}-p_{4}\right)_\mu \nonumber\\
&&\times   \left[\frac{\delta_{a_{2}a_{4}} \delta_{b_{1}b_{4}} \delta_{b_{2}b_{3}}} {(\theta_{14}+\pi {\rm i})(\theta_{23}+\pi {\rm i})(\theta_{24}+\pi {\rm i})}\left(\delta_{a_{0}a_{3}}\delta_{a_{1}c_{0}}-\frac{1}{N}\delta_{a_{0}c_{0}}\delta_{a_{1}a_{3}}\right)\right.\nonumber\\
&&+\frac{\delta_{a_{2}a_{3}}\delta_{b_{1} b_{3}}\delta_{b_{2} b_{4}}}{(\theta_{13}+\pi {\rm i})(\theta_{23}+\pi {\rm i})(\theta_{24}+\pi {\rm i})}\left(\delta_{a_{0} a_{4}}\delta_{a_{1} c_{0}}
-\frac{1}{N}\delta_{a_{0}c_{0}}\delta_{a_{1}a_{4}} \right)\nonumber\\
&&+\frac{\delta_{a_{1} a_{4}}\delta_{b_{1}b_{3}}\delta_{b_{2}b_{4}}}{(\theta_{14}+\pi {\rm i})(\theta_{13}+\pi {\rm i})(\theta_{24}+\pi {\rm i})}\left(\delta_{a_{0} a_{3}}\delta_{a_{2} c_{0}}-\frac{1}{N}\delta_{a_{0} c_{0}}\delta_{a_{2} a_{3}}\right)\nonumber\\
&&\left.+\frac{\delta_{a_{1}a_{3}} \delta_{b_{1} b_{4}}\delta_{b_{2} b_{3}}}{(\theta_{14}+\pi {\rm i})(\theta_{13}+\pi {\rm i})(\theta_{23}+\pi {\rm i})}\left(\delta_{a_{0} a_{4}}\delta_{a_{2} c_{0}}-\frac{1}{N}\delta_{a_{0} c_{0}}\delta_{a_{2} a_{4}}\right)\right]
e^{-ix\cdot(p_{1}+p_{2}+p_{3}+p_{4})} \nonumber \\
&&+O\left(\frac{1}{N^{2}}\right),\label{fourparticle}
\eeq
where $\theta_{ij}=\theta_{i}-\theta_{j}$. 

To find an exact expression for the correlation function, we need the all the form factors (that is, for all $M$). We introduce the permutation $\sigma\in S_{M+1}$ which takes the set of numbers $0,1,\dots,M$ to $\sigma(0),\sigma(1),\dots,\sigma(M)$, respectively, and the permutation $\tau\in S_{M}$ which takes the set of numbers $1,2,\dots,M$ to $\tau(1),\tau(2),\dots,\tau(M)$, respectively. 

The form factor of the current operator with $2M$ excitations is
\beq
\langle 0\!\!\!&\!\!\!\! \vert \!\!\! &\!\!\!\! j_\mu^L(x)_{a_{0} \,a_{\!\,_{{2M+1}}}} \;\vert \; A,\theta_{1},b_{1},a_{1};\dots\, ;A,\theta_M,b_M,a_M;P,\theta_{M+1},a_{M+1},b_{M+1};\dots\, ;P,\theta_{2M},a_{2M},b_{2M}\rangle_{\rm in}\nonumber\\
&=&\langle 0\vert j_\mu^L(x)_{a_{0} a_{\!\,_{\!\,_{2M+1}}}}\mathfrak{A}^\dagger_A(\theta_{1})_{b_{1}a_{1}}\dots\mathfrak{A}^\dagger_A(\theta_M)_{b_{\!\,_{M}} a_{\!\,_{M}}}
\mathfrak{A}^\dagger_{P}(\theta_{M+1})_{a_{\!\,_{M+1}} b_{\!\,_{M+1}}}\dots\mathfrak{A}^\dagger_{P}(\theta_{2M})_{a_{\!\,_{2M}} b_{\!\,_{2M}}}\vert 0\rangle\nonumber\\
&=&\frac{1}{N^{M-1}}
\left(p_{1}+\cdots +p_{M}-p_{M+1}-\cdots -p_{2M}\right)_{\mu}\;
\sum_{\sigma,\tau} F_{\sigma \tau}(\theta_{1},\dots,\theta_{2M})\;  e^{-ix\cdot \sum_{j=1}^{2M}p_{j}}   \nonumber\\
&\times\!\!\!&\!\!\!\left[\prod_{j=0}^M\delta_{a_{j} a_{\sigma(j)+M}}\prod_{k=1}^{M}\delta_{b_{k} b_{\tau(k)+M}}-\frac{1}{N}\delta_{a_{\!\,_{0}} a_{\!\,_{{2M+1}}}}
\delta_{a_{\!\,_{l_{\sigma}}} 
a_{\sigma(0)+M}}\prod_{j=1,\,j\neq l_{\sigma}}^M \delta_{a_{j} a_{\sigma(j)+M}}\prod_{k=1}^M\delta_{b_{k} b_{\tau(k)+M}}\right]
,\label{generalformfactor}
\eeq
where $l_{\sigma}$ is defined by $\sigma(l_{\sigma})+M=2M+1$. This is the most general expression consistent with Lorentz invariance, a traceless current (guaranteed by the second term in square brackets) and crossing. 

To simplify our terminology, we say that excitation $h$ is the particle or antiparticle with rapidity $\theta_{h}$ and left and right indices $a_{h}, b_{h}$, respectively. 

We expand the functions $F_{\sigma \tau}(\theta_{1},\dots,\theta_{2M})$ in powers of $1/N$:
\beq
F_{\sigma \tau}(\theta_{1},\dots,\theta_{2M})=
F^0_{\sigma \tau}(\theta_{1},\dots,\theta_{2M})+\frac{1}{N}F^1_{\sigma \tau}(\theta_{1},\dots,\theta_{2M})+
\frac{1}{N^{2}}F^{2}_{\sigma\tau}(\theta_{1},\dots,\theta_{2M})+\cdots\nonumber
\eeq
keeping only the first term.

The scattering axiom \cite{Smirnov}, also known as Watson's theorem, follows from the Zamolodchikov algebra (\ref{zamoalgebra}) applied to the creation operators in (\ref{generalformfactor}). This axiom implies 
\beq
\langle 0\!\!\!\!\!&\vert&\!\!\! \!\! j_\mu^L(x)_{a_{0} a_{\!\,_{2M+1}}} \mathfrak{A}^\dagger_{I_{1}}(\theta_{1})_{C_{1}}\dots\mathfrak{A}^\dagger_{I_{i}}(\theta_{i})_{C_{i}}\mathfrak{A}^\dagger_{I_{i+1}}(\theta_{i+1})_{C_{i+1}}\dots\mathfrak{A}^\dagger_{I_{2M}}(\theta_{2M})_{C_{2M}}\vert 0\rangle\nonumber\\
&=&S_{I_{i+1} I_{i}}(\theta_{i}-\theta_{i+1})^{C^{\prime}_{i+1};C^{\prime}_{i}}_{C_{i};C_{i+1}}\langle 0\vert j_\mu^L(x)_{a_{0} a_{\!\,_{2M+1}}} \mathfrak{A}^\dagger_{I_{1}}(\theta_{1})_{C_{1}}\dots\mathfrak{A}^\dagger_{I_{i+1}}(\theta_{i+1})_{C^{\prime}_{i+1}}
\mathfrak{A}^\dagger_{I_{i}}(\theta_{i})_{C^{\prime}_{i}}\dots\mathfrak{A}^\dagger_{I_{2M}}(\theta_{2M})_{C_{2M}}\vert 0\rangle,\label{watsonstheorem}
\eeq
where, for each $k$, $I_{k}=P$ for a particle or $I_{k}=A$ for an antiparticle, and $C_{k}$ is the ordered set of indices $C_{k}=(a_{k} ,b_{k})$ for $I_{k}=P$, or $C_{k}=(b_{k}, a_{k})$ for $I_{k}=A$.

We use (\ref{watsonstheorem}) to interchange the creation operator of the excitation $h$ with the creation operator of the excitation $i$ in (\ref{generalformfactor}). 
There are four different ways the function $F^0_{\sigma \tau}(\theta_{1},\dots,\theta_{2M})$ can be affected by interchanging the excitations  $h$ and $i$, for a given $\sigma$ and $\tau$ \cite{Orland2}. If excitation $h$ and excitation $i$ are both particles or both antiparticles, then the rapidities $\theta_{h}$ and $\theta_{i}$ are interchanged in the function $F^0_{\sigma \tau}(\theta_{1},\dots,\theta_{2M})$. If excitation $h$ is a particle, excitation $i$ is an antiparticle, and $\sigma(i)+M\neq h, \tau(i)+M\neq h$, then the function $F^0_{\sigma \tau}(\theta_{1},\dots,\theta_{2M})$ is unchanged. If 
excitation $h$ is a particle, excitation $i$ an antiparticle, and either 
$\sigma(i)+M=h$, $\tau(i)+M\neq h$, or $\sigma(i)+M\neq h$, $\tau(i)+M=h$, then we multiply 
$F^0_{\sigma \tau}(\theta_{1},\dots,\theta_{2M})$ by the pure phase
$\frac{\theta_{ih}+\pi {\rm i}}{\theta_{ih}-\pi {\rm i}}$. If excitation $h$ is a particle, excitation $i$ is an antiparticle and $\sigma(i)+M=h$, $\tau(i)+M=h$, then we multiply the function 
$F^0_{\sigma \tau}(\theta_{1},\dots,\theta_{2M})$ by the pure phase $\left(\frac{\theta_{ih}+\pi {\rm i}}{\theta_{ih}-\pi {\rm i}}\right)^{2}$. 

The rules for interchanging creation operators described in the previous paragraph suggest an underlying Abelian structure for the large-$N$ limit. The pure phase we use in the scattering axiom, namely 
$1$, $\frac{\theta_{ih}+\pi {\rm i}}{\theta_{ih}-\pi {\rm i}}$ or $\left(\frac{\theta_{ih}+\pi {\rm i}}{\theta_{ih}-\pi {\rm i}}\right)^{2}$ is similar to the S-matrix element of 
a theory of colorless particles.

Smirnov's periodicity axiom \cite{Smirnov} states
\beq
\langle 0\vert j_\mu^L(x)_{a_{\!\,_{0}} a_{{\!\,_{2M+1}}}}\mathfrak{A}^\dagger_{I_{1}}\!\!\!\!\!&(&\!\!\!\!\!\theta_{1})_{C_{1}}\mathfrak{A}^\dagger_{I_{1}}(\theta_{2})_{C_{2}}\dots 
\mathfrak{A}^\dagger_{I_M}(\theta_M)_{C_M}\vert 0\rangle\nonumber\\
&=&\langle 0\vert j_\mu^L(x)_{a_{\!\,_{0}} a_{\!\,_{{2M+1}}}}\mathfrak{A}^\dagger_{I_M}(\theta_M-2\pi {\rm i})_{C_M}\mathfrak{A}^\dagger_{I_{1}}(\theta_{1})_{C_{1}}\dots 
\mathfrak{A}^\dagger_{I_{M-1}}(\theta_{M-1})_{C_{M-1}}\vert 0\rangle. \label{perio}
\eeq
In terms of the function $F^0_{\sigma \tau}(\theta_{1},\dots,\theta_{2M})$, (\ref{perio}) is
\beq
F^0_{\sigma \tau}(\theta_{1},\dots,\theta_{2M}) = F^0_{\sigma \tau}(\theta_{2M}-2\pi {\rm i},\theta_{1},\dots,\theta_{2M-1})
=F^0_{\sigma\tau}(\theta_{2M-1}-2\pi {\rm i},\theta_{2M}-2\pi {\rm i},\theta_{1},\dots,\theta_{2M-2})=\cdots .\label{periodicityaxiom}
\eeq
The general solution of (\ref{watsonstheorem}) and (\ref{periodicityaxiom}) is
\beq
F^0_{\sigma\tau}(\theta_{1},\dots,\theta_{2M})=\frac{H_{\sigma \tau}(\theta_{1},\dots,\theta_{2M})}{\prod_{j=1; j\neq l_{\sigma}}^M\left(\theta_{j}-\theta_{\sigma(j)+M}+\pi {\rm i}\right)\prod_{k=1}^M\left(\theta_{k}-\theta_{\tau(k)+M}+\pi {\rm i}\right)}\;,\label{solutionform}
\eeq
where $\sigma(l_{\sigma})+M=2M+1$, and the functions $H_{\sigma \tau}(\theta_{1},\dots,\theta_{2M})$ are holomorphic and periodic in $\theta_{j}$, with
period $2\pi {\rm i}$, for each $j=1,\dots,2M$.

The annihilation-pole axiom \cite{Smirnov} states that
\beq
\langle 0\!\!\!&\vert&\!\!\!j_\mu^L(0)_{a_{0} a_{\!\,_{2M+3}}}\left[\prod_{j=1}^M\mathfrak{A}^\dagger_A(\theta_{j})_{b_{j}a_{j}}\right]\left[\prod_{k=M+1}^{2M}\mathfrak{A}^\dagger_P(\theta_{k})_{a_{k}b_{k}}\right]\mathfrak{A}^\dagger_A(\theta_{{2M+1}})_{b_{\!\,_{2M+1}}a_{\!\,_{2M+1}}}\mathfrak{A}^\dagger_{P}(\theta_{\!\,_{2M+2}})_{a_{\!\,_{2M+2}}b_{\!\,_{2M+2}}}\vert 0\rangle\nonumber\\
&=&[(p_{1}+\cdots+p_M-p_{M+1}-\cdots-p_{2M})+(p_{2M+1}-p_{{2M+2}})]_\mu\;\mathcal{F}(\theta_{1},\dots,\theta_{\!\,_{2M+2}})_{a_{0}a_{2}\dots a_{\!\,_{2M+3}};b_{1}\dots b_{\!\,_{2M+2}}},\nonumber
\eeq
has  a pole at $\theta_{{2M+1}}-\theta_{{2M+2}}=-\pi {\rm i}$, with a residue proportional to the form 
factor of $2M$ excitations. 
\beq
{\rm Res}\!\!\!\!\!\!&&\!\!\! 
{\mathcal F}(\theta_{1},\dots,\theta_{\!\,_{2M+2}})_{a_{0}\dots a_{\!\,_{2M+3}};\; b_{1}\dots b_{\!\,_{2M+2}}}\nonumber\\
&=&2i \mathcal{F}(\theta_{1},\dots,\theta_{2M})_{a_{0} a^{\prime}_{1} \dots a^{\prime}_{2M}a_{\!\,_{2M+3}};\;  b^{\prime}_{1}\;\cdots \;b^{\prime}_{2M}}
\delta_{a^{\prime}_{\!\,_{2M+1}}a_{\!\,_{2M+2}}}
\delta_{b^{\prime}_{\!\,_{2M+1}}b_{\!\,_{2M+2}}}\nonumber\\
&& \times \left[\delta_{a^{\prime}_{1} a_{1}}\delta_{b^{\prime}_{1} b_{1}}
\;\cdots\;
\delta_{a^{\prime}_{\!\,_{2M+1}}a_{\!\,_{2M+1}}}\delta_{b^{\prime}_{\!\,_{2M+1}}b_{\!\,_{2M+1}}}-S_{AA}(\theta_{1\,2M+1})^{b^{\prime}_{\!\,_{2M+1}}
a^{\prime}_{\!\,_{2M+1}};b^{\prime}_{1} a^{\prime}_{1}}_{d_{1}c_{1};b_{1}a_{1}} \;\cdots \;S_{AA}(\theta_{M\,2M+1})^{d_{M-1}c_{M-1};b^{\prime}_{M}   
a^{\prime}_{M}}_{d_Mc_M;b_Ma_M}\right.\nonumber\\
&& \left.\times S_{AP}(\theta_{2M+1\,M+1})^{d_Mc_M;a^{\prime}_{M+1}b^{\prime}_{M+1}}_{c_{M+1}d_{M+1};a_{M+1}b_{M+1}}\;\cdots \;S_{AP}(\theta_{2M+1\,2M})^{c_{2M-1}d_{2M-1};a^{\prime}_{2M}b^{\prime}_{2M}}_{c_{2M}d_{2M};a_{2M}b_{2M}}\right]\,.     \label{annihilationpole}
\eeq

By (\ref{annihilationpole}) the functions $H_{\sigma \tau}(\theta_{1},\dots,\theta_{2M})$ in Equation (\ref{solutionform}) have no singularities with nonzero residues. The minimal choice of each  $H_{\sigma \tau}(\theta_{1},\dots,\theta_{2M})$ is a constant $H_{\sigma \tau}$. The annihilation-pole axiom 
fixes this constant to
\beq
H_{\sigma \tau}=\left\{ \begin{array}{ccc}2\pi {\rm i}(4\pi)^{M-1} &,&\,\,{\rm if\;}\sigma(j)\neq\tau(j),\,{\rm for\,all}\, j   \\
0&,&{\rm otherwise}\end{array}\right.  \;\;. \label{Hconstant}
\eeq
This concludes our derivation of 
all the form factors of the current operator. They are completely specified in (\ref{generalformfactor}), (\ref{solutionform}) and (\ref{Hconstant}).

\section{Vacuum Expectation Values of Products of Current Operators }

The current-current correlation function is
\beq
W^{j}_{\mu\nu}(x)_{a_{0} c_{0} ;e_{0}  f_{0}}=\langle 0 \vert j^{L}_{\mu}(x)_{a_{0}c_{0}} \; j^{L}_{\nu}(0)_{e_{0}f_{0}} \vert 0 \rangle
=\sum_{M=1}^{\infty} W_{\mu\nu}^{2M}(x)_{a_{0}c_{0}e_{0}f_{0}}, \label{currcorr}
\eeq
where the contribution from the $2M$-excitation form factor is given by
\beq
W_{\mu\nu}^{2M}(x)_{a_{0}c_{0}e_{0}f_{0}}&=&\frac{1}{N(M!)^{2}}\int\frac{d\theta_{1}\dots d\theta_{2M}}{(2\pi)^{2M}}
\;e^{-ix\cdot\sum_{j=1}^{2M} p_{j} }\nonumber\\
&&\times \langle 0 \vert j_\mu^L(0)_{a_{0}c_{0}}\vert A,\theta_{1},b_{1},a_{1};\dots;A,\theta_M,b_Ma_M;P,\theta_{M+1},a_{M+1},b_{M+1};\dots;P,\theta_{2M},a_{2M},b_{2M}\rangle_{\rm in}\nonumber\\
&&\times\langle0\vert j_\nu^L(0)_{e_{0}f_{0}}\vert  A,\theta_{1},b_{1},a_{1};\dots;A,\theta_M,b_Ma_M;P,\theta_{M+1},a_{M+1},b_{M+1};\dots;P,\theta_{2M},a_{2M},b_{2M}\rangle_{\rm in}^*.\nonumber
\eeq
Substituting the form factors (\ref{generalformfactor}), (\ref{solutionform}), (\ref{Hconstant}), we find
\beq
W_{\mu\nu}^{2M}(x)_{a_{0}c_{0}e_{0}f_{0}}&=&\frac{1}{(M!)^{2}}\int \prod_{j=1}^{2M}\frac{d\theta_{j}}{4\pi} \;\;
e^{-ix\cdot \sum_{j=1}^{2M}p_{j}}    \nonumber\\
&&\times
(p_{1}+\cdots+p_M-p_{M+1}-\cdots-p_{2M})_\mu
(p_{1}+\cdots+p_M-p_{M+1}-\cdots-p_{2M})_\nu      \nonumber\\
&&\times\left[\sum_{\sigma,\tau\in S_M}\frac{\vert H_{\sigma \tau}\vert^{2}    (\delta_{a_{0}e_{0}}\delta_{c_{0}f_{0}}-\frac{1}{N}\delta_{a_{0}c_{0}}\delta_{e_{0}f_{0}})}
{\prod_{j=1;\,j\neq l_{\sigma}}^{M}\vert\theta_{j}-\theta_{\sigma(j)+M}+\pi {\rm i}\vert^{2}\prod_{k=1}^{M}
\vert \theta_{k}-\theta_{\tau(k)+M}+\pi {\rm i} \vert^{2}}+\mathcal{O}\left(\frac{1}{N^{2}}\right)\right],\label{wightmanone}
\eeq
where we have used
\beq
\sum_{a_{1},\dots,a_{2M},b_{1},\dots,b_{2M}}\left[\prod_{j=0}^M\delta_{a_{j}a_{\sigma(j)+M}}\prod_{k=1}^M\delta_{b_{k}b_{\tau(k)+M}}\right.&-&\left.\frac{1}{N}\delta_{a_{0}a_{\!\,_{2M+1}}}\delta_{a_{l_{\sigma}}a_{\sigma(0)+M}}\prod_{j=1;\,j\neq l_{\sigma}}^M\delta_{a_{j}a_{\sigma(j)+M}}\prod_{k=1}^M\delta_{b_{k}b_{\tau(k)+M}}\right]\nonumber\\
\times\left[\prod_{j=0}^M\delta_{a^{\prime}_{j} a^{\prime}_{\omega(j)+M}}\prod_{k=1}^M\delta_{b_{k}b_{\varphi(k)+M}}\right.&-&\left.\frac{1}{N}\delta_{a^{\prime}_{0} 
a^{\prime}_{\!\,_{2M+1}}}\delta_{a^{\prime}_{l_\omega}a^{\prime}_{\omega(0)}}\prod_{j=1\,j\neq l_{\omega}}^M\delta_{a^{\prime}_{j} a^{\prime}_{\omega(j)+M}}\prod_{k=1}^M\delta_{b_{k}b_{\varphi(k)+M}}\right]\nonumber\\ \nonumber \\
=N^{2M-1}
(\,
 \delta_{a_{0} e_{0}}\delta_{c_{0} f_{0}} 
&-& 
\delta_{a_{0}c_{0}}\delta_{e_{0} f_{0}}/N \,)
\left[ \delta_{\sigma\omega}\delta_{\tau\varphi}+\mathcal{O}\left(\frac{1}{N^{2}}\right) \right],\nonumber
\eeq
where $\{a_{j} \}=a_{0},a_{1},a_{2},\dots,a_{2M},c_{0}$ and $\{a^{\prime}_{j}\}=e_{0},a_{1},a_{2},
\dots,a_{2M},f_{0}$. 

The contribution to (\ref{wightmanone}) from each pair $\sigma,\tau$ is the same (because there is no change if the integration variables are interchanged). There are $(M!)^{2}$ pairs $\sigma,\tau$ that satisfy $H_{\sigma \tau}\neq0$, by (\ref{Hconstant}). We choose the contribution from one pair $\sigma,\tau$ in 
(\ref{wightmanone}) and multiply it by $(M!)^{2}$. We choose $\tau(j)=j,$ for $j=1,\dots,M$, and 
$\sigma(1)=2M+1, \sigma(j)=j-1,$ for $j=2,\dots,M,$ such that
\beq
W_{\mu\nu}^{2M}(x)_{a_{0}c_{0}e_{0}f_{0}}&=&\int
\prod_{j=1}^{2M}\frac{d\theta_{j}}{4\pi}
\;\;e^{-ix\cdot\sum_{j=1}^{2M}p_{j}} 
\;4\pi^{2}(4\pi)^{2M-2}\;
\;\left(\delta_{a_{0}e_{0}}\delta_{c_{0}f_{0}}-\frac{1}{N}\delta_{a_{0}c_{0}}\delta_{e_{0}f_{0}}\right)
 \nonumber\\
&\times&
(p_{1}+\cdots+p_M-p_{M+1}-\cdots-p_{2M})_\mu
(p_{1}+\cdots+p_M-p_{M+1}-\cdots -p_{2M})_\nu
\nonumber\\
&&\times  \frac{1}{(\theta_{1}-\theta_{M+1})^{2}+\pi^{2}} \frac{1}{(\theta_{2}-\theta_{M+2})^{2}+\pi^{2}} \dots \frac{1}{(\theta_M-\theta_{2M})^{2}+\pi^{2}}  
\nonumber\\
&&\times    \frac{1}{(\theta_{2}-\theta_{M+1})^{2}+\pi^{2}} \frac{1}{(\theta_{3}-\theta_{M+2})^{2}+\pi^{2}} \dots\ \frac{1}{(\theta_M-\theta_{2M-1})^{2}+\pi^{2}}
+O\left(\frac{1}{N^{2}}\right)\nonumber
\eeq
We further relabel the integration variables as $\theta_{1}\to\theta_{1}, \theta_{2}\to\theta_{3}, \theta_{3}\to\theta_5,\dots,\theta_M\to\theta_{2M-1};\theta_{M+1}\to\theta_{2}, \theta_{M+2}\to\theta_{4},\dots,\theta_{2M}\to\theta_{2M}$. This yields the expression for the non-time-ordered correlation function of two current operators:
\beq
W_{\mu\nu}^{j}(x)_{a_{0}c_{0}e_{0}f_{0}}
&=&\left(\delta_{a_{0}e_{0}}\delta_{c_{0}f_{0}}-
\frac{1}{N}\delta_{a_{0}c_{0}}\delta_{e_{0}f_{0}}\right)
\nonumber\\
&\times &\sum_{M=1}^{\infty}\frac{1}{4}\int  \prod_{j=1}^{2M}d\theta_{j}\;\;
\;e^{-ix\cdot \sum_{j=1}^{2M}p_{j} }\;
\;P^{M}_{\mu} P^{M}_{\nu}
\prod_{j=1}^{2M-1}
\frac{1}{(\theta_{j}-\theta_{j+1})^{2}+\pi^{2}}\;+\;O\left(\frac{1}{N^{2}}\right),   \label{finalMcurr}
\eeq
where
\beq
P^{M}=\sum_{j=1}^{2M} (-1)^{j}p_{j}
 \;.\label{Pdef}
\eeq

\section{Form Factors of the Stress-Energy-Momentum Tensor}

There is usually some ambiguity in the definition of the stress-energy-momentum 
operator \cite{Coleman}. Some examples of this ambiguity, in the context of the 
form-factor program, have been examined by Mussardo and Simonetti \cite{MussardoSimonetti}. As mentioned in the introduction, the only ambiguity for the principal chiral model is the 
coefficient of the cosmological-constant contribution. In our case, we use Smirnov's axioms, in particular the ``minimality axiom" (that is, the form factors are as nonsingular as possible) and local Lorentz invariance as 
a guide to a proper definition of the stress-energy-momentum tensor. This does not prove that we have made the correct choice. On the other hand, we are confident that this is the case. As is pointed out in Reference 
\cite{MussardoSimonetti}, different field theories (represented by different renormalization-group fixed points) can have the same S matrix, but different correlation functions. Mussardo and Simonetti showed that the ambiguity can be parametrized by the matrix element of the trace of the stress-energy-momentum tensor between a one-excitation state and the vacuum:
\beq
F_{1}=\langle 0\vert T^{\mu}_{\mu}(0)\vert \theta\rangle_{\rm in}\; \nonumber
\eeq
(we have not explicitly written the colors of the ket, nor specified whether it is a particle or antiparticle). In our case, however, this quantity must be fixed to zero. This is simply because the $(1+1)$-dimensional vacuum state and the trace
have no (particle number or color) quantum numbers.

This stress-energy-momentum tensor operator is invariant under ${\rm SU}(N)\!\times \!{\rm SU}(N)$ transformations. Thus the only non-vanishing form factors have equal number of particles and antiparticles in the in-state ket. The general form factor with $M$ particles and $M$ antiparticles is
\beq
\langle 0\vert\;T_{\mu\nu}(0)\!\!\!&\!\!\!\vert\!\!\!&\!\!\!A,\theta_{1},b_{1},a_{1};\dots;A,\theta_M,b_M,a_M;P,\theta_{M+1},a_{M+1},b_{M+1};\dots;P,\theta_{2M},a_{2M},b_{2M}\rangle_{\rm in}\nonumber\\
&=&\left(p_{1}+\cdots+p_M-p_{M+1}-\cdots-p_{2M}\right)_\mu\left(p_{1}+\cdots+p_M-p_{M+1}-\cdots-p_{2M}\right)_\nu\nonumber\\
&&\times\frac{1}{N^{M-1}}\sum_{\sigma,\tau\in S_M}F_{\sigma\tau}(\theta_{1},\dots,\theta_{2M})\prod_{j=1}^M\delta_{a_{j}a_{\sigma(j)+M}}\prod_{k=1}^M\delta_{b_{k}b_{\tau(k)+M}},\label{stressformfactor}
\eeq
where $\sigma, \tau\in S_M$ are permutations of the integers $1,2,\dots,M$ (this is different from the convention in Section III. Recall that there the permutation $\sigma$ was defined as an element of $S_{M+1}$).

We expand the function 
$F_{\sigma,\tau}(\theta_{1},\dots,\theta_{2M})$ in powers of $1/N$, {\em i.e.}, as $F^0_{\sigma,\tau}(\theta_{1},\dots,\theta_{2M})+\frac{1}{N}F^1_{\sigma,\tau}(\theta_{1},\dots,\theta_{2M})+\cdots$, keeping only the first term.

The form factors in (\ref{stressformfactor}) behave the same way as the current-operator form factors under Watson's theorem and the periodicity axiom. These two axioms give us the solution
\beq
F^0_{\sigma\tau}(\theta_{1},\dots,\theta_{2M})=\frac{H_{\sigma \tau}(\theta_{1},\dots,\theta_{2M})}{\prod_{j=1}^M(\theta_{j}-\theta_{\sigma(j)+M}+\pi {\rm i})\prod_{k=1}^M(\theta_{k}-\theta_{\tau(k)+M}+\pi {\rm i})}.\label{SEMFF}
\eeq
The minimal choice is to make $H_{\sigma,\tau}(\theta_{1},\dots,\theta_{2M})=H_{\sigma \tau}$ constants. These constants can be fixed by the annihilation pole axiom, once we fix the constant for the two-particle form factor.

For $M=1$, Equation (\ref{stressformfactor}) becomes
\beq
\langle 0\vert T_{\mu\nu}(x)\vert  A,\theta_{1},b_{1},a_{1};P,\theta_{2},a_{2},b_{2}\rangle_{\rm in}=(p_{1}-p_{2})_\mu(p_{1}-p_{2})_\nu \frac{g}{(\theta_{12}+\pi {\rm i})^{2}}
e^{-ix(p_{1}+p_{2})}\delta_{a_{1}a_{2}}\delta_{b_{1}b_{2}}+O\left(\frac{1}{N}\right).\label{twoparticlestress}
\eeq
We fix the constant $g$ by requiring that 
\beq
\int\,dx^1 \,T_{00}(x)\vert  A,\theta_{1},b_{1},a_{1}\rangle_{\rm in}=m\cosh\theta_{1}\vert  A,\theta_{1},b_{1},a_{1}\rangle_{\rm in}.\label{stressconstantfixing}
\eeq
Notice that the pole in (\ref{twoparticlestress}) has vanishing residue. Therefore, by the annihilation-pole axiom, the vacuum energy
is zero. 

We next apply crossing, changing one of the incoming particles in (\ref{twoparticlestress}) to an outgoing antiparticle, and integrate over the spatial coordinate $x^{1}$, yielding
\beq
\int \,dx^1\;_{\rm in}\langle A,\theta_{2},b_{2},a_{2}\vert T_{00}(x)\vert \!\!\!&\!\!\!A\!\!\!&\!\!\!,\theta_{1},b_{1},a_{1}\rangle_{\rm in} \nonumber\\
&=& (m\cosh\theta_{1}+m\cosh\theta_{2})^{2} \; 2\pi 
\delta(m\sinh\theta_{1}-m\sinh\theta_{2})
\frac{ g}{(\theta_{12}+2\pi {\rm i})^{2}}\delta_{a_{1}a_{2}}\delta_{b_{1}b_{2}}+O\left(\frac{1}{N}\right)
\nonumber\\
&=&- \frac{m g}{2\pi^{2}}\cosh\theta_{1}  \;  4\pi \delta(\theta_{12}) \;\delta_{a_{1}a_{2}}  \delta_{b_{1}b_{2}}
+O\left(\frac{1}{N}\right).
\nonumber
\eeq
The condition (\ref{stressconstantfixing}) implies $g=-2\pi^{2}$.

The constants $H_{\sigma \tau}$ for the $2M$-particle form factor are fixed by the annihilation pole axiom (Equation (\ref{annihilationpole})), which gives the values
\beq
H_{\sigma \tau}=\left\{\begin{array}{ccc}(-2\pi^{2})(4\pi)^{M-1}\!\!\!&,&{\rm for}\,\sigma(j)\neq\tau(j),\,{\rm for\,all}\,j\\
0\;&,&\!\!\!\!\!\!\!\!\!\!{\rm otherwise}\end{array}\right.   \;. \label{SEMH}
\eeq
The $2M$-particle form factor has a total of $(M!)^{2}/2$ non-vanishing terms.

To summarize the results of this section, (\ref{stressformfactor}), (\ref{SEMFF}), (\ref{SEMH}) determine all the form factors of the stress-energy-momentum tensor.

\section{The Correlation Function of the Stress-Energy-Momentum Tensor }

In this section we obtain the vacuum expectation value of the product of two stress-energy-momentum-tensor 
operators. In other words, we find
\beq
W^{T}_{\mu\nu\alpha\beta}(x)=\frac{1}{N^{2}}\langle 0\vert T_{\mu\nu}(x)T_{\alpha\beta}(0)\vert 0\rangle
=\sum_{M=1}^\infty W_{\mu\nu\alpha\beta}^{2M}(x)\,,\label{SEMTsum}
\eeq
where the terms in the sum over $M$ are defined as
\beq
W_{\mu\nu\alpha\beta}^{2M}(x)&=&\frac{1}{N^{2}}\frac{1}{(M!)^{2}}\int\prod_{j=1}^{2M}\frac{d\theta_{j}}{4\pi}
\;e^{-ix\cdot\sum_{j=1}^{2M}p_{j}}\nonumber\\
&&\times\langle 0\vert T_{\mu\nu}(0)
\vert A,\theta_{1},b_{1},a_{1};\dots;A,\theta_M,b_M,a_M;P,\theta_{M+1},a_{M+1},b_{M+1};\dots;P,\theta_{2M},a_{2M},b_{2M}\rangle_{\rm in}.\nonumber\\
&&\times\langle 0\vert T_{\alpha\beta}(0)\vert  A,\theta_{1},b_{1},a_{1};\dots;A,\theta_M,b_M,a_M;P,\theta_{M+1},a_{M+1},b_{M+1};\dots;P,\theta_{2M},a_{2M},b_{2M}\rangle_{\rm in}^*. \nonumber
\eeq

Substituting our form factors (\ref{stressformfactor}), (\ref{SEMFF}), (\ref{SEMH}) gives
\beq
W_{\mu\nu\alpha\beta}^{2M}(x)&=&\frac{1}{(M!)^{2}}\int
\prod_{j=1}^{2M}\frac{d\theta_{j}}{4\pi}
\;e^{-ix\cdot\sum_{j=1}^{2M}p_{j}}\nonumber\\
&&\times \left(p_{1}+\cdots+p_M-p_{M+1}-\cdots-p_{2M}\right)_\mu\left(p_{1}+\cdots+p_M-p_{M+1}-\cdots-p_{2M}\right)_\nu\nonumber\\
&&\times\left(p_{1}+\cdots+p_M-p_{M+1}-\cdots-p_{2M}\right)_\alpha\left(p_{1}+\cdots+p_M-p_{M+1}-\cdots-p_{2M}\right)_\beta\nonumber\\
&&\times\sum_{\sigma,\tau\in S_M}\frac{\vert H_{\sigma \tau}\vert^{2}}{\prod_{j=1}^M\left(\theta_{j}-\theta_{\sigma(j)+M}+\pi {\rm i}\right)^{2}\prod_{k=1}^{M}\left(\theta_{k}-\theta_{\tau(k)+M}+\pi {\rm i}\right)^{2}}+\mathcal{O}\left(\frac{1}{N}\right),\nonumber
\eeq
where we have used
\beq
\sum_{a_{1},\dots,a_{2M},b_{1},\dots,b_{2M}}\left[\prod_{j=1}^M\prod_{k=1}^M\delta_{a_{j}a_{\sigma(j)+M}}\delta_{b_{k}b_{\tau(k)+M}}\right]\left[\prod_{j=1}^M\prod_{k=1}^M\delta_{a_{j}a_{\omega(j)+M}}\delta_{b_{k}b_{\varphi(k)+M}}\right]=N^{2M}\left[\delta_{\sigma\omega}\delta_{\tau\varphi}+\mathcal{O}\left(\frac{1}{N}\right)\right].\nonumber
\eeq

The contribution to $W_{\mu\nu\alpha\beta}^{2M}(x)$ from each pair $\sigma,\tau$ is the same. There are $\frac{(M!)^{2}}{2}$ possible  pairs $\sigma,\tau$. We write the contribution from just one of these pairs, and multiply by the factor $\frac{(M!)^{2}}{2}$. We choose $\sigma(j)=j$ for $j=1,\dots,M$, and $\tau(1)=M,\,\tau(j)=j-1$ for $j=2,\dots,M$. Then we have
\beq
W_{\mu\nu\alpha\beta}^{2M}(x)&=&\frac{1}{2}\int
\prod_{j=1}^{2M}\frac{d\theta_{j}}{4\pi}\;
e^{-ix\cdot\sum_{j=1}^{2M}p_{j}}4\pi^4(4\pi)^{2M-2}\nonumber\\
&&\times \left(p_{1}+\cdots+p_M-p_{M+1}-\cdots-p_{2M}\right)_\mu\left(p_{1}+\cdots+p_M-p_{M+1}-\cdots-p_{2M}\right)_\nu\nonumber\\
&&\times\left(p_{1}+\cdots+p_M-p_{M+1}-\cdots-p_{2M}\right)_\alpha\left(p_{1}+\cdots+p_M-p_{M+1}-\cdots-p_{2M}\right)_\beta\nonumber\\
&&\times
\frac{1}{\left(\theta_{1}-\theta_{M+1}\right)^{2}+\pi^{2}} \frac{1}{\left(\theta_{2}-\theta_{M+2}\right)^{2}+\pi^{2}} \cdots 
\frac{1}{\left(\theta_M-\theta_{2M}\right)^{2}+\pi^{2}}\nonumber\\
&&\times\frac{1}{\left(\theta_{1}-\theta_{2M}\right)^{2}+\pi^{2}} 
\frac{1}{\left(\theta_{2}-\theta_{M+1}\right)^{2}+\pi^{2}}
\cdots  \frac{1}{\left(\theta_M-\theta_{2M-1}\right)^{2}+\pi^{2}}+O\left(\frac{1}{N}\right).
\label{SEMCF}
\eeq
Finally, as in Section IV, we relabel the integration variables by $\theta_{1}\to\theta_{1},\;\theta_{2}\to\theta_{3},\;\theta_{3}\to\theta_5,\;\dots,\;\theta_M\to\theta_{2M-1},\;\theta_{M+1}\to\theta_{2},\;
\theta_{M+2}\to\theta_{4},\;\dots,\;\theta_{2M}\to\theta_{2M}$. This gives the expression for the non-time-ordered correlation function
\beq
W^{T}_{\mu\nu\alpha\beta}(x)&=&\frac{\pi^{2}}{8}\sum_{M=1}^{\infty}\int\ \prod_{j=1}^{2M}d\theta_{j}\;
e^{-ix\cdot\sum_{j=1}^{2M}p_{j}}
\; P^{M}_{\mu}P^{M}_{\nu}P^{M}_{\alpha}P^{M}_{\beta}\nonumber\\
&&\times
\frac{1}{(\theta_{1}-\theta_{2M})^{2}+\pi^{2}}\prod_{j=1}^{2M-1}\frac{1}{(\theta_{j}-\theta_{j+1})^{2}+\pi^{2}}
+O\left(\frac{1}{N}\right)
\;,
\label{SEMCF2}
\eeq
where the vector $P^{M}$ is given by (\ref{Pdef}). 

\section{The Abelian nature of form factors in the 't$\;$Hooft limit}

At large-$N$, the application of the 2-body particle-antiparticle S matrix with relative rapidity $\theta$ is equivalent to multiplying by a pure phase. As discussed in Section III, the phase is 
$(\theta+\pi{\rm i})/(\theta-\pi{\rm i})$ raised to the power of the number of 
contracted color indices. 

The form factors at $N=\infty$ (we must drop the terms of order $1/N$ for the current form factors) contain 
Kronecker deltas in color indices. In the case of operators such as the renormalized field or the current, the
Kronecker deltas can be represented diagrammatically as lines. This is shown in Figure 1. Any set of 
contractions topologically different than that shown in this figure is of higher order in $1/N$. This indicates 
a natural ordering for the incoming excitations. We have labeled these $1,2,\dots,n$. We emphasize that
this integer index is not necessarily the same as the subscript of rapidities or colors. A similar structure exists
for color singlet operators, such as the stress-energy-momentum tensor; this is shown in Figure 2. The
only feature different from Figure 1 is that there is a color contraction between the first and last excitations. It is significant that particles and antiparticles alternate in these figures. 

The excitation index $j=1,\dots,n$, shown in Figures 1 and 2, inspires the notion of ``nearest neighbor"
excitations. Nearest-neighbor pairs are indexed by $j$ and $j+1$ and possibly $n$ and $1$ (for a color singlet operator). The odd indices correspond to incoming particles (antiparticles) and
the even indices correspond to incoming antiparticles (particles). The main point is that
each excitation can scatter nontrivially {\em only} with its nearest 
neighbors. We stress that this property appears to be related to the 
't$\,$Hooft limit of amplitudes, rather than integrability. 

We now make the planarity property described above more explicit. Let us introduce
excitation creation operators ${\mathfrak A}^{\dagger}(\theta)_{j}$, where $j=1,\dots, n$. Two such operators commute
unless they correspond to nearest-neighbor excitations. We replace the original Zamolodchikov algebra 
(\ref{zamoalgebra}) by
\beq
{\mathfrak A}^{\dagger}(\theta)_{j}\,{\mathfrak A}^{\dagger}(\theta^{\prime})_{k}
&=&\frac{\theta^{\prime}-\theta+\pi{\rm i}}{\theta^{\prime}-\theta-\pi{\rm i}}
\;{\mathfrak A}^{\dagger}(\theta^{\prime})_{k}\,{\mathfrak A}^{\dagger}(\theta)_{j},
\;\; k=j+1, \nonumber \\
{\mathfrak A}^{\dagger}(\theta)_{j}\,{\mathfrak A}^{\dagger}(\theta^{\prime})_{k}
&=&\frac{\theta^{\prime}-\theta-\pi{\rm i}}{\theta^{\prime}-\theta+\pi{\rm i}}
\;{\mathfrak A}^{\dagger}(\theta^{\prime})_{k}\,{\mathfrak A}^{\dagger}(\theta)_{j}
,\;\; k=j-1, \nonumber \\ 
{\mathfrak A}^{\dagger}(\theta)_{j}\,{\mathfrak A}^{\dagger}(\theta^{\prime})_{k}
&=&
{\mathfrak A}^{\dagger}(\theta^{\prime})_{k}\,{\mathfrak A}^{\dagger}(\theta)_{j}
\;,\;\;\;\;\;\;\;\;\;\;\;\;\;\;\;\;\;\;\;\; {\rm otherwise}. \label{AbelZamo}
\eeq
In the case of a color singlet operator, we define the addition operation to be modulo $n$, that is
$n+1=1$, $1-1=n$ in (\ref{AbelZamo}). This algebra is associative. The associativity is trivial, and
does not appear to be related to integrability at finite $N$. This leads us to ask the pregnant
question: can the form-factor bootstrap work for the 't$\;$Hooft limit of a non-integrable field theory?


\section{Conclusions}

To summarize, we have obtained all the form factors of the current vector and stress-energy-momentum tensor 
in the SU($N$) principal chiral model as $N\rightarrow \infty$. We have used this 
to find vacuum expectation values of products of these operators. Together with the result for the 
Wightman function of the scaling field \cite{Orland2}, this brings us closer to a complete picture of this quantum field theory.

As discussed in Section VII, the simple nature of form factors at large $N$ may not require integrability
at finite $N$. The prospect of a bootstrap program 
for the planar limit of non-integrable field theories is exciting and deserves to be explored further. 

The induced Yang-Mills action and induced gravitational action can be determined from the correlation functions of the currents and the stress-energy momentum tensor, respectively. Our result cannot completely fully determine these effective actions, as we only have correlation functions of two operators. Higher-point correlation functions are harder to determine, because there are ``disconnected" pieces in form-factor expansions of these functions, which must be subtracted  \cite{BabujianKarowEtc.}. We believe that this obstacle is 
surmountable, however.

Our work should have application in the study of $(2+1)$-dimensional SU($N$) gauge theories as coupled 
$(1+1)$-dimensional principal chiral models \cite{2+1}, and perhaps in other problems with SU($N$) symmetry.

\begin{acknowledgments}
We would like to thank Robert Konik for discussions. P.O. thanks the staff of the Niels Bohr Institute for their hospitality while some of this work was done. A.C.C.'s research was supported in part by the Weissman School of Arts and Sciences at 
Baruch College, and by a Dean K. Harrison Award. P.O.'s research was supported in 
part by the National Science Foundation, under Grant No. 
PHY0855387, and by a grant from the PSC-CUNY. 
\end{acknowledgments}

\newpage

\begin{figure}

\begin{picture}(150,75)(30,0)

\put(90, 40){\oval(100, 60)[tl]}
\put(90,85){\oval(10,30)[br]}
\put(105,85){\oval(10,30)[bl]}
\put(150, 40){\oval(40, 60)[tr]}

\put(50,40){\oval(10,40)[t]}
\put(65,40){\oval(10,40)[t]}
\put(80,40){\oval(10,40)[t]}

\put(160,40){\oval(10,40)[t]}

\put(110,70){\circle*{2}}
\put(125,70){\circle*{2}}
\put(140,70){\circle*{2}}

\put(110,50){\circle*{2}}
\put(125,50){\circle*{2}}
\put(140,50){\circle*{2}}

\put(41,32){\large\bf 1}
\put(56,32){\large\bf 2}
\put(71,32){\large\bf 3}
\put(167,32){\large\bf n}

\end{picture}

\caption{Illustration of the index structure of a form factor of an operator with two color indices (such as a local field or current), as $N\rightarrow\infty$. Lines denote contractions of excitation indices. In this figure and the next, the integers $1,2,\dots,n$ denote the order of the excitations, according to how indices are contracted. One color index from each of the first and last excitations is contracted with the operator. The excitations alternate between particles and
antiparticles. }

\end{figure}
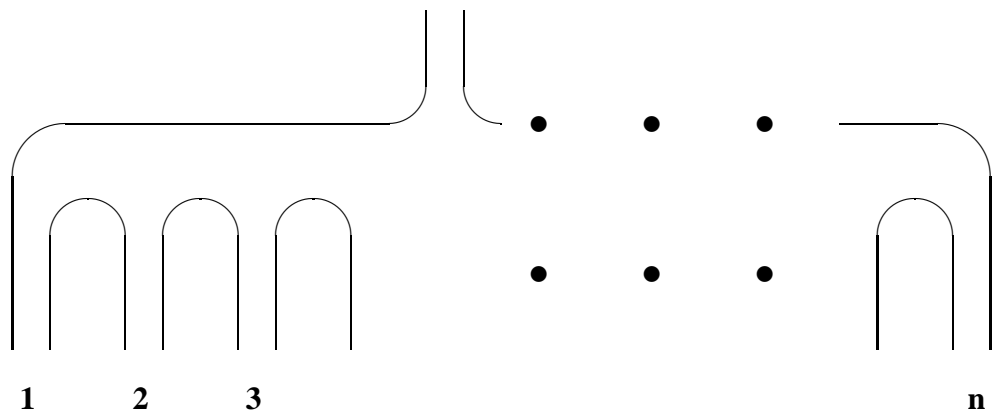

\hfill

\newpage

\begin{figure}

\begin{picture}(150,75)(30,0)

\put(100, 40){\oval(100, 60)[tl]}
\put(150, 40){\oval(40, 60)[tr]}

\put(60,40){\oval(10,40)[t]}
\put(75,40){\oval(10,40)[t]}
\put(90,40){\oval(10,40)[t]}
\put(160,40){\oval(10,40)[t]}

\put(110,70){\circle*{2}}
\put(125,70){\circle*{2}}
\put(140,70){\circle*{2}}

\put(110,50){\circle*{2}}
\put(125,50){\circle*{2}}
\put(140,50){\circle*{2}}

\put(51,32){\large\bf 1}
\put(67,32){\large\bf 2}
\put(81,32){\large\bf 3}
\put(166,32){\large\bf n}

\end{picture}

\caption{Illustration of the index structure of the form factor of a color-singlet operator (such as the stress-energy-momentum tensor) as $N\rightarrow\infty$. In this case, a color index of the first excitation is contracted with a color index of the last excitation.}

\end{figure}
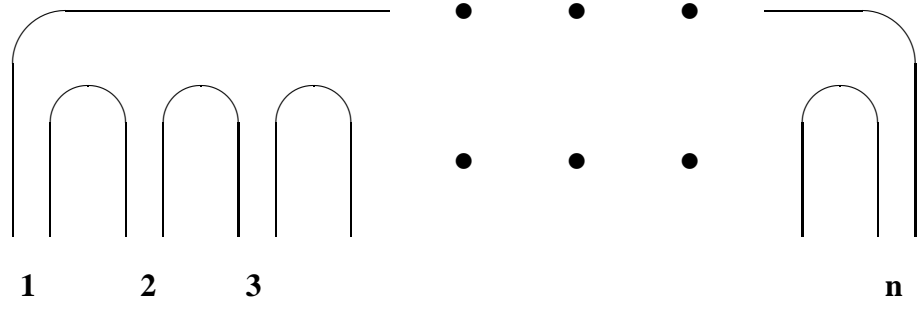

\end{document}